\documentstyle[prl,epsfig,aps,multicol]{revtex}  

\title{Closed Path Integrals and Renormalisation in Quantum Mechanics}

\author{H.~Jirari$^{a}$, 
H.~Kr\"{o}ger$^{a}\footnote{Corresponding author, Email: hkroger@phy.ulaval.ca}$, 
X.Q.~Luo$^{b}$, 
K.J.M.~Moriarty$^{c}$
and S.G.~Rubin$^{d}$ }

\address{$^{a}$D\'{e}partement de Physique, Universit\'{e} Laval, Qu\'{e}bec,
  Qu\'{e}bec G1K 7P4, Canada \\ 
  $^{b}$Department of Physics, Zhongshan University, 
  Guangzhou 510275, China \\
  $^{c}$Department of Mathematics, Statistics and Computer Science,
  Dalhousie University, Halifax, Nova Scotia B3H 3J5, Canada \\
  $^{d}$Moscow Engineering Physics Institute, Center for Cosmo-Particle Physics 
 "Cosmion", Moscow, Russia }
      
\date{\today} 

\begin{document} 

\newcommand{\co}[1]{{\bf[#1]}}
\newcommand{\be}{\begin{equation}}
\newcommand{\ee}{\end{equation}}
\newcommand{\bea}{\begin{eqnarray}}
\newcommand{\eea}{\end{eqnarray}}
\newcommand{\sgn}{\mbox{sgn}}
\newcommand{\slesssim}{{\scriptstyle \lesssim}}

\renewcommand{\vec}[1]{{\bf #1}}

\maketitle

\begin{abstract}
We suggest a closed form expression for the path integral 
of quantum transition amplitudes. 
We introduce a quantum action with renormalized parameters. 
We present numerical results for the $V \sim x^{4}$ potential.
The renormalized action is relevant for quantum chaos and quantum instantons.
\end{abstract}

\pacs{Valid PACS appear here}

\begin{multicols}{2}

\noindent {\bf 1. Introduction} \\ 
The path integral has become a standard method to quantize classical theories.
The class of systems
for which the path integral
can be computed
analytically
is ridiculously
small.
Examples are quadratic Lagrangians, e.g., for free motion and 
the harmonic oscillator \cite{Schulman:81}.
Let us consider the Q.M. transition amplitude from 
$x_{in}$, $t_{in}$ to $x_{fi}$, $t_{fi}$ given by 
\be
\label{PathIntegral}
G(x_{fi},t_{fi};x_{in},t_{in}) =
\left. \int [dx] \exp[ \frac{i}{\hbar} 
S[x] ] \right|_{x_{in},t_{in}}^{x_{fi},t_{fi}}  ,
\ee
where $S = \int dt  \frac{m}{2} \dot{x}^{2} - V(x)$ 
denotes the classical action.
In some cases this path integral can be expressed as a sum over 
classical paths only
\be
\label{SumClassPath}
G(x_{fi},t_{fi}; x_{in},t_{in}) = 
\sum_{ \{x_{cl}\} } Z \exp \left[ \frac{i}{\hbar} 
\left. S[{x}_{cl}] \right|_{x_{in},t_{in}}^{x_{fi},t_{fi}} \right] ,
\ee
where $S[x_{cl}]$ is the classical action evaluated along the 
classical trajectory from $x_{in}$, $t_{in}$ to $x_{fi}$, $t_{fi}$. 
There may be infinitely many classical paths for a given pair of boundary points
$(x_{in}, t_{in})$, $(x_{fi}, t_{fi})$ \cite{Schulman:81,Schulman:87}.   
Different paths then correspond to different values of the action.
The factor $Z$ represents some (time-dependent) normalisation factor, 
enforced by the unitarity of the amplitude. 
Eq.(\ref{SumClassPath}) holds, e.g., for quadratic Lagrangians, 
where the sum runs over a single path where the action is minimal.
In the case of the harmonic oscillator $V(x) = \frac{m}{2} \omega^{2} x^{2}$,
one has 
\begin{eqnarray*}
\label{HarmonicOsc}
&& \left. S[x_{cl}] \right|_{x_{in},t_{in}}^{x_{fi},t_{fi}} 
 = \frac{ m \omega}{2 \sin(\omega T) } 
\left[ (x_{fi}^{2} + x_{in}^{2}) \cos(\omega T) - 2 x_{in}x_{fi} \right] , 
\nonumber \\
&& Z =  \sqrt{ \frac { m \omega }{ 2 \pi i \hbar \sin(\omega T) } }, ~~~
T = t_{fi}-t_{in} . 
\end{eqnarray*}
Another example where the sum over classical paths is exact is the 
path integral of the quantum mechanical top, which mathematically 
corresponds to free motion on the group manifold of $SU(2)$ \cite{Schulman:68}.
Eq.(\ref{SumClassPath}) is a well known but remarkable result.
It is possible that there are other systems, where the path integral is 
given by a sum over classical paths. However, the result is not true in general. A simple counter example is given by Nelson \cite{Nelson}.

In the following we will explore if the validity of Eq.(\ref{SumClassPath}) 
can be extended to a wider class of quantum systems, if we allow the 
classical action to be replaced by a renormalized action. 
The motivation comes from perturbative renormalisation in Q.F.T.
Recall that a theory is called renormalizable, if a fixed finite 
number of counter terms of the action allows to make the Q.F.T. finite to any given order.
The renormalized action is different from the bare action, due to quantum 
loop corrections, which exist for interacting theories. The action and also Green's functions 
have the same structure in its bare and renormalized form.
Quantum mechanics can be viewed as Q.F.T. in $0+1$ dimension. 
Here we suggest for Q.M. the existence of a renormalized/quantum action, 
which describes the transition amplitude (Green's function), having the same 
structure but parameters different from the classical action.

{\it Conjecture}:
For a given classical action $S = \int dt \frac{m}{2} \dot{x}^{2} - V(x)$ 
there is a renormalized action 
$\tilde{S} = \int dt \frac{\tilde{m}}{2} \dot{x}^{2} - \tilde{V}(x)$, 
which allows to express the transition amplitude by
\be
\label{DefRenormAction}
G(x_{fi},t_{fi}; x_{in},t_{in}) = \tilde{Z} 
\exp [ \frac{i}{\hbar} \left. \tilde{S}[\tilde{x}_{cl}] 
\right|_{x_{in},t_{in}}^{x_{fi},t_{fi}} ] .
\ee
Here $\tilde{x}_ {cl}$ denotes the classical path corresponding to 
the action $\tilde{S}$, such that the action $\tilde{S}(\tilde{x}_{cl})$ 
is minimal (we exclude the occurrence of conjugate points or caustics). 
$\tilde{Z}$ denotes the normalisation factor corresponding to $\tilde{S}$. 
Eq.(\ref{DefRenormAction}) is valid with 
the {\em same} action $\tilde{S}$ for all sets of 
boundary positions $x_{fi}$, $x_{in}$ for a given time interval $T=t_{fi}-t_{in}$. 
The parameters of the renormalized action depend on the time $T$. The 
renormalized action converges to a non-trivial limit when $T \to \infty$. 
Any dependence on $x_{fi}, x_{in}$ enters via the trajectory 
$\tilde{x}_ {cl}$. $\tilde{Z}$ depends on the action parameters and $T$, 
but not on $x_{fi}, x_{in}$.

Why is such result of physical interest?
(i) Having constructed the action $\tilde{S}$
gives a closed form solution for the path integral. 
(ii) The action $\tilde{S}$ defines a renormalized action in Q.M. 
(iii) Gutzwiller's trace formula \cite{Gutzwiller:90}
has been used widely                                                                 to study quantum chaos in a semi-classical regime         
(e.g. highly excited states in atoms). 
Eq.(\ref{DefRenormAction})      
establishes a simple relation between
the transition amplitude and some action $\tilde{S}$ which
mathematically has the form of the classical action.
Eq.(\ref{DefRenormAction})
allows to define quantum chaos in an unambigous way.
(iv) Eq.(\ref{DefRenormAction}) allows to study quantum descendents
of classical instanton solutions.
We have no proof of this conjecture. However, we will give numerical
evidence in support of the conjecture. We will elaborate
on the topic (ii), while (iii,iv) will be discussed in another letter.

\bigskip

\noindent {\bf 2. Construction of renormalized action} \\
We know the quantum action for the harmonic oscillator. What does perturbation theory predict for the quantum action in the presence of a small anharmonic perturbation $S[x] = \int_{0}^{T} ~ \frac{m}{2} \dot{x}^{2} + \frac{m \omega^{2}}{2} x^{2} + \lambda V(x)$, $\lambda << 1$? 
Using the saddle point method one obtains up to order $O(\lambda^{2})$
\begin{eqnarray*}
\tilde{S} &=& S + S_1 + S_2,
\nonumber \\
S_1[x] &=& \frac{\lambda}{2m} \int_{0}^{T} dt ~ G(t,t) ~ V''(x(t)) ,
\nonumber \\
S_2[x] &=& \frac{\lambda^{2}}{4m^{2}} 
\int_{0}^{T} dt dt'  G(t,t')  G(t',t)  V''(x(t))  V''(x(t')) ,
\nonumber \\
G(t,t') &=& \frac{1}{\omega sh(\omega T)} 
\left[ 
\begin{array}{c}
sh(\omega t) ~ sh(\omega (t'-T)): t \leq t' \\
sh(\omega t') ~ sh(\omega (t-T)): t' \leq t 
\end{array}
\right] ,
\end{eqnarray*}
where $G(t,t')$ denotes the Euclidean Green's function. This shows that the quantum action contains terms beyond that occuring in the classical action.

In the following we want to determine numerically the renormalized 
action, Eq.(\ref{DefRenormAction}), for the case of a quartic potential. 
In this work, we search for renormalized parameters, neglecting potential terms higher than fourth order in a first step.
This search requires to calculate the 
transition amplitude, Eq.(\ref{PathIntegral}).
We have chosen to compute the path integral via Monte Carlo 
with importance sampling. This requires to go over to 
imaginary time (Euclidean path integral) $t \to -i t$. 
Then Eq.(\ref{PathIntegral}) becomes
\begin{eqnarray*}
G_{E}(x_{fi},t_{fi};x_{in},t_{in}) =
\left. \int [dx] \exp[ - \frac{1}{\hbar} 
S_{E}[x] ] \right|_{x_{in},t_{in}}^{x_{fi},t_{fi}}
\end{eqnarray*}
where the Euclidean classical action is given by
$S_{E} = \int dt  \frac{m}{2} \dot{x}^{2} + V(x)$. 
Correspondingly, Eq.(\ref{DefRenormAction}) becomes
\be
\label{DefEuclRenormAction}
G_{E}(x_{fi},t_{fi}; x_{in},t_{in}) = \tilde{Z}_{E} 
\exp \left[ - \frac{1}{\hbar} \left. 
\tilde{S}_{E}[\tilde{x}_{E}^ {cl} ] \right|_{x_{in},t_{in}}^{x_{fi},t_{fi}} \right] , 
\ee
where the Euclidean quantum action is given by
$\tilde{S}_{E} = \int dt \frac{\tilde{m}}{2} \dot{x}^{2} 
+ \tilde{V}(x)$.
The transition to imaginary time causes a sign change of the mass term, 
besides that the action parameters keep its absolute and relative values. 
In the sequel of the paper we will work in imaginary time.
We drop the subscript $E$ in what follows.

We consider a particle of mass $m$ moving in the presence of a local potential 
$V(x)$, given in polynomial form
$V(x) = \sum_{n=0}^{N} v_{n} x^{n}$, with 
the parameters $m, v_{0},\dots,v_{N}$.
We write the potential of the quantum action 
$\tilde{V}(x) = \sum_{n=0}^{N} \tilde{v}_{n} x^{n}$, 
being parametrized by 
$\tilde{m}, \tilde{v}_{0},\dots,\tilde{v}_{N}$. 
We choose a value for the time $T$ 
and a set of boundary points $\{x_1,\dots, x_J\}$.
For all pairs of $x_{in}$, $x_{fi}$ from that set
we compute the Euclidean transition amplitude $G_{ij} \equiv G(x_{i},x_{j},T)$. 
We denote the Euclidean action $\tilde{S}$ along its classical path 
$\tilde{x}_ {cl}$ by
$\tilde{\Sigma}_{ij} = \tilde{S}[\tilde{x}_{cl} ] |_{x_{j},0}^{x_{i},T}$.
Because $\tilde{Z}$ does not depend on $x_{in}$, $x_{fi}$,
we subsume it into 
$\tilde{\Sigma}^{sub}_{ij} = \tilde{\Sigma}_{ij} - \ln\tilde{Z}$.
Then Eq.(\ref{DefEuclRenormAction}) takes the form
\be
\label{SystEqs}
G_{ij} =  \exp[ - \tilde{\Sigma}^{sub}_{ij} ], ~~~ i,j = 1,...,J .
\ee
While the quantum action has $N+2$ parameters, Eq.(\ref{SystEqs})
represents $J^{2}/2 + O(J)$ independent equations to determine the quantum action. 
We have chosen a larger number of equations than parameters (over-determined) 
requiring that the error in Eq.(\ref{SystEqs}) becomes globally minimal. 
I.e., we make a $\chi^{2}$-fit,
\begin{eqnarray*}
\chi^{2} = \sum_{i,j=1,..,J} 
( G_{ij} - \exp[ - \tilde{\Sigma}^{sub}_{ij} ])^{2} 
/\sigma_{ij}^{2} ,
\end{eqnarray*}
which is a function of the parameters of the quantum action. 
A solution for the quantum action is considered as consistent,
if we find the same action parameters for different sets of boundary points.

\bigskip

\noindent {\bf 3. Numerical results} \\
We have computed the Euclidean propagator in two ways: 
(i) We use Monte Carlo with importance sampling. The trick is 
to write the propagator matrix element as a ratio of two path integrals. 
This corresponds to splitting the action $S= S_{0} + S_{1}$, 
chosen such that the path integral for $S_{0}$ is analytically known, 
$\exp[-S_{0}/\hbar]$ is treated as weight and $\exp[-S_{1}/\hbar]$ 
is treated as observable. (ii) According to the Feynman-Kac formula, 
the propagator for large time is asymptotically given by the ground 
state contribution. We have solved the stationary Schr\"odinger equation,    
computing a number $M$ (of lowest lying) energies and wave functions, 
and expressed the propagator as a spectral sum over those states.

(a) To test our algorithms, we considered the harmonic oscillator
$V(x) = v_{2} x^{2}, ~ v_{2}=1/2$ (with parameters $m=\omega=\hbar=1$). 
The classical path, the transition amplitude and the renormalized action 
are analytically known, giving $\tilde{S}=S$.
The results are found within statistical errors to be consistent with $\tilde{S}=S$.  

(b) Next we have considered the quartic interaction 
$V(x) = v_{4} x^4, ~ v_{4}=1$, with parameters $m=\hbar=1$.
We have computed numerically the ground state energy, $E_{gr}=0.667986$, 
and the variance of the ground state wave function, $var_{gr}=0.287333$.
From those we introduce dynamically a time scale and a length scale 
(corresponding to the Bohr radius),
$T_{sc}= \hbar/E_{gr} = 1.497$, $L_{sc} = \sqrt{var_{gr}} = 0.5360$. 
We have computed the propagator for $0 < T <2$  
via Monte Carlo runs with $N_{therm}=4000$, $N_{skip}=2000$ and $N_{conf}=1000$. 
For $T > 1$ we also computed the propagator by solving the stationary 
Schr\"odinger equation and calculating the wave function and energy 
of the lowest $M$  states of the spectrum. We used $M=30$ for $1 < T < 2$ and 
$M=7$ for $T > 2$. We used up to 20 000 mesh points to solve the classical 
equation of motion in the determination of $\tilde{S}[\tilde{x}_{cl}]$.
For the time interval $T=0.5$, the numerical results are shown in Tab.[1]. 
The following observations are made:
(i) The linear and the cubic renormalized potential term are compatible 
within statistical error with the value zero,  
expected from parity conservation of the quantum system. 

\end{multicols}

Tab.1. Quartic potential $V(x) = x^{4}$. Harmonic oscillator weight factor. $T=0.5$, $J=6$. \\
\begin{tabular}{|l|l|l|l|l|l|l|l|l|} \hline \hline
& $\tilde{m}$ & $\tilde{v_{0}}$ & $\tilde{v_{1}}$ & $\tilde{v_{2}}$ & $\tilde{v_{3}}$ & $\tilde{v_{4}}$ & $\chi^{2}$ &  \\ \hline
Fit MC & 0.9936(3) & 1.1711(18) & 0.000(7) & 0.449(18) & 0.000(16) & 0.982(23) & 21.35 & [-1.0,+1.0] \\ \hline 
Fit MC & 0.9938(2) & 1.1665(19) & 0.000(7) & 0.488(15) & 0.000(13) & 0.954(15) & 17.47 & [-1.2,+1.2] \\ \hline 
Fit MC & 0.9941(2) & 1.1695(20) & 0.000(7) & 0.466(12) & 0.000(11) & 0.975(10) & 24.98 & [-1.4,+1.4] \\ \hline 
Fit MC & 0.9940(2) & 1.1696(21) & 0.000(8) & 0.469(10) & 0.000(10) & 0.973(7) & 63.01 & [-1.6,+1.6] \\ \hline 
Fit MC & 0.9944(2) & 1.1684(23) & 0.000(8) & 0.458(9) & 0.000(9) & 0.984(5) & 38.82 & [-1.8,+1.8] \\ \hline 
Fit MC & 0.9942(2) & 1.1679(25) & 0.000(9) & 0.459(9) & 0.000(9) & 0.986(4) & 55.73 & [-2.0,+2.0] \\ \hline 
Fit MC & 0.9939(2) & 1.1721(26) & 0.000(9) & 0.449(8) & 0.000(8) & 0.992(4) & 57.80 & [-2.2,+2.2] \\ \hline 
Fit MC & 0.9938(2) & 1.1689(27) & 0.000(10) & 0.460(9) & 0.000(8) & 0.987(3) & 43.45 & [-2.4,+2.4] \\ \hline 
Fit MC & 0.9942(3) & 1.1664(29) & 0.000(11) & 0.447(9) & 0.000(8) & 0.993(3) & 55.34 & [-2.6,+2.6] \\ \hline
Fit MC & 0.9938(3) & 1.1738(29) & 0.000(12) & 0.432(10) & 0.000(8) & 0.998(3) & 58.76 & [-2.8,+2.8] \\ \hline 
Fit MC & 0.9946(3) & 1.1705(31) & 0.000(13) & 0.435(10) & 0.000(9) & 0.990(3) & 44.48 & [-3.0,+3.0] \\ \hline 
Fit MC & 0.9940(4) & 1.1702(34) & 0.000(13) & 0.461(11) & 0.000(9) & 0.984(3) & 64.68 & [-3.2,+3.2] \\ \hline 
Fit MC & 0.9943(4) & 1.1645(37) & 0.000(15) & 0.460(11) & 0.000(10) & 0.985(4) & 63.00 & [-3.4,+3.4] \\ \hline 
Fit MC & 0.9949(5) & 1.1605(41) & 0.000(16) & 0.483(12) & 0.000(11) & 0.973(4) & 47.03 & [-3.6,+3.6] \\ \hline \hline
Average & 0.9941(3) & 1.1685(27) & 0.000(10) & 0.458(11) & 0.000(10) & 0.983(7) & 46.85 &  \\ \hline \hline 
\end{tabular} 
\\

\begin{multicols}{2} 

(ii) The renormalized action generates a quadratic potential term, 
absent in the classical action. 
With respect to variation of the interval, most coefficients 
vary very little and are compatible within statistical errors  with each other.
The quadratic term is more sensitive, showing fluctuations from its 
average value of up to two standard deviations. In general we find for 
$J=6$ that the numerical values of $\chi^{2}$ lie in the order of the 
value $\chi^{2}=36$, which means that the fits are acceptable. 
We found that the intervals, where the algorithms work, lie in the 
range of $[-1.0,+1.0]$ to $[-3.6,+3.6]$. 
If the interval is too small, with boundary points $x_{i} << 1$, then 
the potential terms $x^{2}$ and $x^{4}$ are difficult to distinguish.
On the other hand, if the interval is too large, with boundary points 
$x_{i}>>1$, then the observable $\exp[-x^{4}]$ (of the Monte Carlo algorithm) 
becomes very small. In the renormalized action the quartic term dominates 
strongly over the quadratic term, which makes the latter difficult to discriminate. Stability of the
renormalized parameters under variation of the interval 
of boundary points has been observed also for other values of the 
time $T$. The dependence on $T$ is shown in Fig.[1]. The error bars are estimated systematical errors (statistical errors of 
Monte Carlo data are of the size of symbols).
We distinguish three regimes: (i) $0 < T << T_{sc}$,
(ii) $T \approx T_{sc}$, and (iii) $T >> T_{sc}$. For $T$ close to  
zero, the renormalized parameters are close to those of the classical action.
An exception is $\tilde{v}_{4}$ at $T=0.1$, which is about $10\%$ off 
the classical value. We believe that this does not reflect physics but is a 
fault due to limited numerical precision (recall that the propagator becomes 
$\delta(x_{fi}-x_{in})$ when $T \to 0$).   
At about $T \approx T_{sc}$, $\tilde{m}$, $\tilde{v}_{2}$ and $\tilde{v}_{4}$ 
change noticeably. For large $T >> T_{sc}$ the renormalized parameters 
converge asymptotically. When increasing $T$ beyond $T_{sc}$ one observes 
that the stability of results requires an exponential increase in the 
number of mesh points. This can be traced to the fact that the classical 
trajectory covers several orders of magnitude (and the order increases 
with $T$). This observation, similar to critical slowing down observed 
in simulating critical phenomena in lattice field theory, 
puts an upper limit on the time parameter used in this investigation ($T \leq 5$).

\bigskip

\noindent {\bf 4. Interpretation} \\
How can we understand the behavior of the renormalized parameters? 
Why does the renormalized action 
depend on $T$?  
Consider the n-point function of a scalar field, 
\begin{eqnarray*} 
\Gamma = <vac| t.o.(\Phi(x_{1}) \cdots \Phi(x_{n}) |vac> .
\end{eqnarray*}
This is the vacuum expectation value of a time-ordered n-fold product 
of the field. It corresponds to the transition from the physical vacuum at 
$t= -\infty$, to the physical vacuum at $t=+\infty$ with intermediate 
creation (annihilation) of particles at $t_{1}, \cdots, t_{n}$. 
In order to obtain finite expressions, one introduces a regularisation 
parameter, say $\mu$. Then 
$\Gamma = \Gamma(x_{i},\mu)$, i.e. the result depends on $\mu$. 
In Q.M., we consider the propagator from $x_{in}, t_{in}$ to 
$x_{fi}, t_{fi}$, $T=t_{fi}-t_{in}$. The analogue in Q.F.T. is 
\begin{eqnarray*} 
<vac,t=T | t.o.(\Phi(x_{1}) \cdots \Phi(x_{n}) |vac, t=0 > .
\end{eqnarray*}
Then the n-point function $\Gamma(x_{i},\mu,T)$ depends also on the time $T$. 
The computation of renormalized parameters from $\Gamma(x_{i},\mu,T)$
then gives mass, coupling constants etc. as function of $T$.
Thus it is no surprise that the renormalized parameters 
in Q.M. also depend on $T$.

\vspace{8.0cm}

\noindent Fig.[1] Renormalized parameters (in dimensionless units) versus time $T$. Transition amplitude obtained by Monte Carlo (dots and circles) and by stationary Schr\"odinger equation (other symbols).  
\\

How can we understand the behavior of the renormalized 
parameters changing qualitatively in the investigated 
time interval?
(i) For sufficiently small $T$, one has $S < \hbar$, i.e. we are in the 
quantum regime. However, according to Dirac \cite{Dirac:33} 
(see also Ref.\cite{Schulman:81}),
$\exp(iS/\hbar)$ is a good approximation of the propagator $G$, 
when the time interval $T$ over which $G$ is supposed to propagate goes to zero. This is consistent with our numerical observation $\tilde{S} \approx S$ in this regime.
(ii) For sufficiently large time $T$ one has $S > \hbar$, i.e. one is in the 
semi-classical (WKB) regime. In imaginary time, the Feynman-Kac formula 
describes the propagator asymptotically,
\begin{eqnarray*}
G(x_{fi},T; x_{in},0) \sim_{T \to \infty}
\psi_{gr}(x_{fi}) \exp(-E_{gr} T/\hbar) \psi_{gr}(x_{in}) . 
\end{eqnarray*}
For the harmonic oscillator ($E_{gr}=\hbar \omega/2$)  
one can show 
\begin{equation}
\label{Asymptote}
\tilde{v}_{0} \sim_{T \to \infty} E_{gr} .
\end{equation}
For the quartic potential one observes also a 
smooth behavior of $\tilde{v}_{0}$. We have fitted $\tilde{v}_{0}$ 
by the function $A + B/T$. We obtain $\tilde{v}_{0} \to 0.6686$, 
compatible with Eq.(\ref{Asymptote}).
Then using the definition of the renormalized action, 
Eq.(\ref{DefEuclRenormAction}), splitting the renormalized action 
into a $\tilde{v}_{0}$ part and a rest, and using the Feynman-Kac formula implies
\begin{eqnarray*}
&& \exp[ - \frac{1}{\hbar} \int_{0}^{T} dt ~ \tilde{v}_{0} ] ~~
\exp[ - \frac{1}{\hbar} \int_{0}^{T} dt \frac{\tilde{m}}{2} 
\dot{\tilde{x}}_{cl}^{2} + \tilde{V}(\tilde{x}_{cl}) ]  
\nonumber \\
&& \longrightarrow_{T \to \infty} 
\psi_{gr}(x_{fi}) \exp(-E_{gr} T/\hbar) \psi_{gr}(x_{in}) . 
\end{eqnarray*}
Consequently, the renormalized action, when excluding the 
$\tilde{v}_{0}$ term, has an asymptotic limit
for large $T$
\begin{eqnarray*}
- \frac{1}{\hbar} \int_{0}^{T} dt \frac{\tilde{m}}{2} 
\dot{\tilde{x}}_{cl}^{2} + \tilde{V}(\tilde{x}_{cl})   
\longrightarrow_{T \to \infty} 
\ln[ \psi_{gr}(x_{fi}) ~ \psi_{gr}(x_{in}) ] . 
\end{eqnarray*}
This is a strong indication (no proof) that the 
renormalized parameters of $\tilde{S}$ converge asymptotically in $T$.
This is what has been observed in the data. It corresponds in Q.F.T. to the infinite volume limit.

In conclusion, we have computed numerically for the quartic potential a quantum corrected action, and find parameters being quite different from the classical values. Hence, quantum fluctuations can change the classical behavior drastically. This has been observed (in preliminary results) also for the instantons solutions of the double well potential.

\bigskip

\noindent {\bf Acknowledgements} \\ 
H.K. and K.M. are grateful for support by NSERC Canada.
H.K. would like to acknowledge discussions with L.S. Schulman.

\end{multicols}

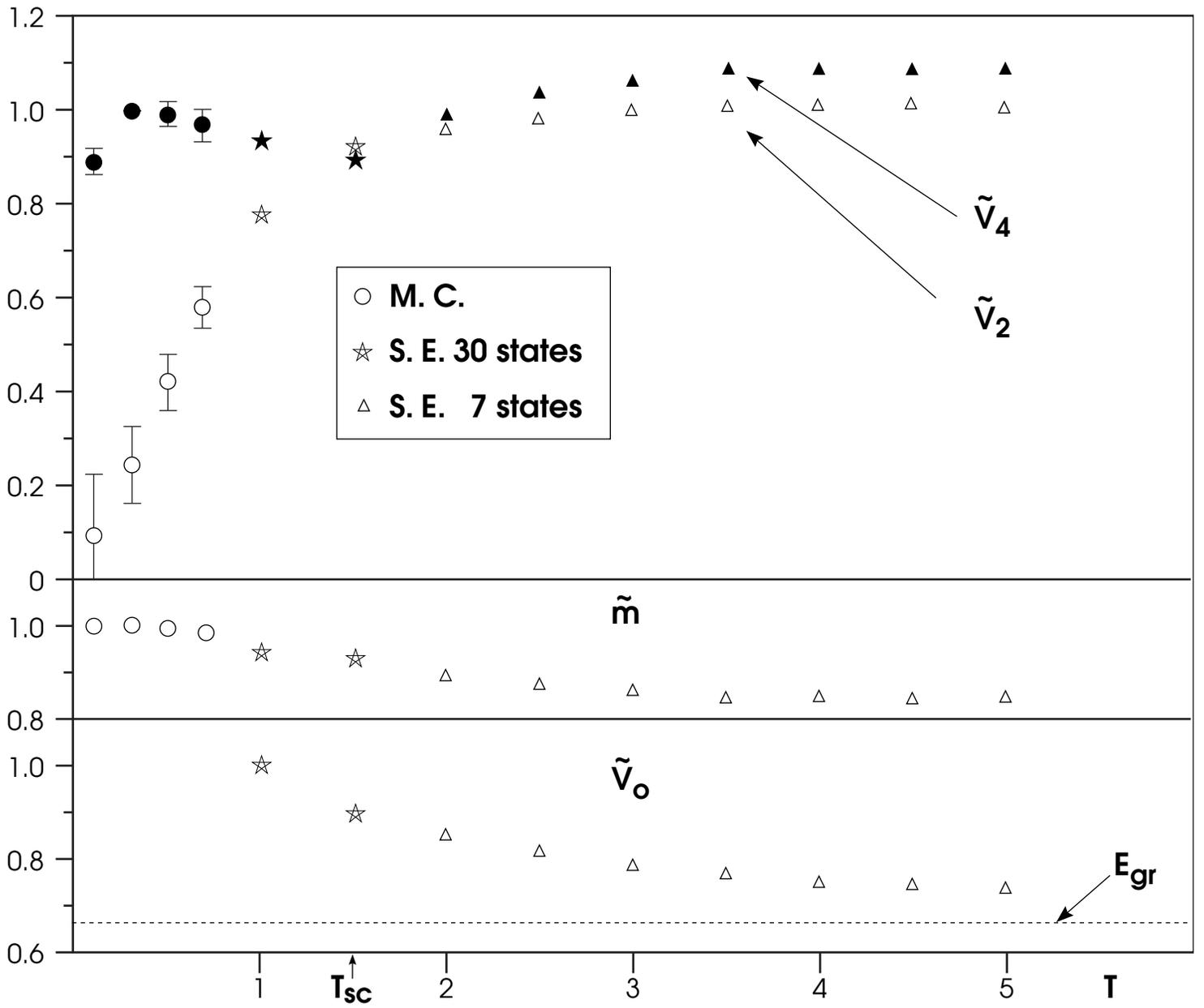
\end{document}